 \documentstyle[aps,multicol]{revtex}
\draft
\begin{document}
\title{Analog of Kelvin-Helmholtz instability on
superfluid liquid free surface}
\author{S. E. Korshunov$~^{1}$ }
\address {{L. D. Landau Institute for Theoretical Physics RAS,}
        {Kosygina 2, 117940 Moscow, Russia}}
\date{April 19, 2002}
\maketitle
\begin{abstract}
We analyse the analog of the Kelvin-Helmholtz instability on free surface
of a superfluid liquid. This instability is induced by the relative motion
of superfluid and normal components of the same liquid along the surface.
The instabilty threshold is found to be independent of the value
of viscosity, but turns out to be lower than in absence of dissipation.
The result is similar to that obtained for the interface between two
sliding superfluids (with different mechanism of dissipation)
and confirmed by the first experimental observation of the Kelvin-Helmholtz
instability on the interface between $^3$He-A and $^3$He-B
by Blaauwgeers {\em et al.} (cond-mat/0111343). \\
\end{abstract}
 % (current)

\begin{multicols}{2}

\section{Introduction}
\footnotetext[1]{email: serkor@landau.ac.ru}

The Kelvin-Helmholtz instability \cite{LL1} is a dynamic corrugation
instability of the interface separating two liquids
sliding with respect to each other.
The concept of such instability has been orginally introduced when
considering ideal liquids and in presence of dissipation becomes ill
defined, because a relative motion of two liquids in contact with each
other is no longer a solution of the hydrodynamic equations.

The simplest situation, when an equilibrium difference in velocities
can be maintained at a surface of a liquid, is the relative motion
of superfluid and normal components (a counterflow) in superfluid
$^4\mbox{He}$.
The corrugation instability of a superfluid liquid free surface
in presence of a counterflow along the surface was studied
in Ref. \cite{K91}
(in relation with experiments of Egolf {\em et al.} \cite{Egolf}).
It can be considered as an example of the Kelvin-Helmholtz instability,
in which both liquids are located on the same side of the interface.
Analogous instability can appear when superfluid $^4\mbox{He}$
is sliding along the atomically-rough interface
separating it from solid $^4\mbox{He}$ \cite{Kagan}.
Such interface is known to allow for equilibrium melting
and crystallization of $^4\mbox{He}$ \cite{Andr,KPB}, and, as  a
consequence, its behavior resembles that of free surface of a liquid.

Recently the interest to surface instabilities in superfliuds has
been revived \cite{MCBD,SC,V} in relation
with experiments on laser manipulated Bose gases
and the first experimental observation of the Kelvin-Helmholtz
instability on the interface between two superfluids, $^3$He-A
and $^3$He-B \cite{BEV}.
In particular, it has been demonstrated \cite{V}
that addition of a friction related to the motion of the interface
with respect to container walls shifts the point of instability from
the well known classical threshold \cite{LL1} to another value.
This value does not depend on the strength of dissipation and can be
reproduced in the framework of the thermodynamic analysis by
looking for the instability of the free energy calculated
in the reference frame of normal component, which in equilibrium
is at rest with respect to container walls.
The appearance of the same threshold in dynamic analysis has been
ascribed in Ref. \cite{V} to the symmetry breaking related with
the violation of the Galilean invariance by the considered friction force.

In the present work we return to investigation of the corrugation
instability on free surface of a supefluid liquid
in presence of a counterflow \cite{K91},
taking into account the viscosity of the normal component,
and show that for any finite value of viscosity the instability
threshold is shifted to viscosity independent value, which is in
agreement with the results of Ref. \cite{V}.
However, in our analysis this phenomenon appears in absence
of the friction force violating the Galilean invariance.
Therefore, the modification of the instability criterion in presence
of dissipation is not a consequence of the symmetry breaking form
of the friction, but has a more general nature.

\section{Dispersion relation}
The calculation of the surface oscillations spectrum in
superfluid liquid in presence of a counterflow
can be performed in the same way
as the calculation of the spectrum of gravitational wave
in normal liquid with finite viscosity \cite{LL2}.
For frequencies small in comparison with that of the first and the second
sound, the mass and the entropy densities can be assumed to be constant.
Accordingly, the conservation laws for mass and entropy are reduced to
constraints:
\begin{equation}
\mbox{div}\,{\bf v}_{\rm s}=\mbox{div}\,{\bf v}_{\rm n}=0\;,  \label{div}
\end{equation}
where ${\bf v}_{\rm s}$ and ${\bf v}_{\rm n}$ are superfluid
and normal velocities.
In that limit the Navier-Stokes equation for superfluid liquid
can be written as \cite{LL3}
\begin{eqnarray}
\rho_{\rm s}\left[\frac{\partial {\bf v}_{\rm s}}{\partial t}
+({\bf v}_{\rm s}\nabla){\bf v}_{\rm s}\right] & + &
\rho_{\rm n}\left[\frac{\partial {\bf v}_{\rm n}}{\partial t}
+({\bf v}_{\rm n}\nabla){\bf v}_{\rm n}\right] = \nonumber \\
 & = & -\nabla p -\rho{\bf g}+\eta\Delta{\bf v}_{\rm n}\;,  \label{NS}
\end{eqnarray}
where $\rho_{\rm s}$ and $\rho_{\rm n}$ are superfluid and normal
densities ($\rho=\rho_{\rm s}+\rho_{\rm n}$ being the total density),
$p$ is the pressure, ${\bf g}$ is the free fall acceleration,
and $\eta$ is the viscosity.

Solution of Eqs. (\ref{div})-(\ref{NS}), satisfying
the constraint $\mbox{rot}\,{\bf v}_{\rm s}=0$ and corresponding to a
small amplitude surface wave with frequency $\omega$ and wavevector
${\bf q}$ parallel to the surface (we assume that in the equilibrium
the liquid is situated at $z<0$), can be chosen in the form
\begin{eqnarray}
{\bf v}_{\rm s}^{\|}({\bf r},t) & = & {\bf v}_{\rm s}^0
+i{\bf q}\gamma e^{qz}A   \;,         \label{b1}\\
v_{\rm s}^{z}({\bf r},t) & = & q\gamma e^{qz}A
\;,              \\
{\bf v}_{\rm n}^{\|}({\bf r},t) & = & {\bf v}_{\rm n}^0+
i{\bf q}\gamma (e^{qz}B+e^{kz}C ) \;,    \label{b3} \\
v_{\rm n}^z({\bf r},t) & =  & \gamma [qe^{qz}B+(q^2/k)e^{kz}C]
\;, \\ p({\bf r},t) & = & -\rho gz+           \label{b5}   \\
 & & +i\gamma e^{qz}                                 \nonumber
 [\rho_{\rm s} (\omega-{\bf v}_{\rm s}^0{\bf q} ) A
+\rho_{\rm n} (\omega-{\bf v}_{\rm n}^0{\bf q} ) B ]\;,
\end{eqnarray}
where
superscript $\|$ refers to the component of a vector
parallel to the surface,
\makebox{$\gamma=\exp i({\bf qr-\omega t})$},
\begin{equation}
k=\sqrt{q^2-i\frac{\rho_{\rm n}}{\eta}
\left(\omega-{\bf v}_{\rm n}^0{\bf q}\right)}~,~~~\mbox{Re}\,k>0\;,
\end{equation}
$A$, $B$ and $C$ are (yet arbitrary) constants,
and the possibility of an equilibrium counterflow
(characterized by ${\bf v}_{\rm s}^0\neq{\bf v}_{\rm n}^0$) is taken into
account.

Substitution of Eqs. (\ref{b1})-(\ref{b5}) into the boundary conditions
describing the conservation of mass and entropy
\begin{equation}
{v}_{\rm s}^z-({\bf v}_{\rm s}^{\|}\nabla^\|)\zeta
={v}_{\rm n}^z-({\bf v}_{\rm n}^{\|}\nabla^\|)\zeta
= \frac{\partial \zeta}{\partial t}\;,
\end{equation}
and mechanical equilibrium
\begin{eqnarray}
\eta\left(\nabla^z {\bf v}_{\rm n}^\|+\nabla^\|{v}_{\rm n}^z\right) & = & 0
\;, \\ -p+2\eta\nabla^z{{v}_{\rm n}^z} & = &
\sigma(\nabla^\|)^2\zeta
\end{eqnarray}
at the surface (whose deviation from the plane $z=0$ is denoted by
$\zeta$ and surface tension by $\sigma$)
shows that they are compatible with each other when
\begin{eqnarray}
\rho_{\rm s}\left(\omega-{\bf v}_{\rm s}^0{\bf q}\right)^2+
\rho_{\rm n}\left(\omega-{\bf v}_{\rm n}^0{\bf q}+i\frac{2\eta
q^2}{\rho_{\rm n}}\right)^2+ \nonumber & & \\
+\frac{4\eta^2q^3k}{\rho_{\rm n}} = \rho gq+\sigma q^3 &. &   \label{dr}
\end{eqnarray}
The derivation of Eq. (\ref{dr}) have not required to assume
the viscosity small, so it is applicable for arbitrary value of viscosity.

\section{Instability thresholds for zero and finite viscosity}

For $\rho_{\rm s}=0$ and $\sigma=0$ Eq. (\ref{dr}) is transformed into
the dispersion relation of gravitational wave on free surface of
normal liquid \cite{LL2}, whereas in the limit of $\eta=0$ it is reduced
to equation
\begin{equation}
\left(\omega-{\bf v}{\bf q}\right)^2=gq+\frac{\sigma}{\rho}q^3
-\frac{\rho_{\rm n}\rho_{\rm s}}{\rho^2}({\bf wq})^2       \label{dr0}
\end{equation}
describing the spectrum of surface wave in superfliud with
a counterflow \cite{K91}
derived in the framework of the non-dissipative two-fluid description.
Here ${\bf v}=(\rho_{\rm s}{\bf v}_{\rm s}^0
+\rho_{\rm n}{\bf v}_{\rm n}^0)/\rho$ is the mass velocity and
${\bf w}={\bf v}_{\rm n}^0-{\bf v}_{\rm s}^0$ the relative
velocity in the superfluid.
The form of Eq. (\ref{dr0}) shows that the roots
with positive and negative imaginary parts
(the former corresponds to growing corrugation) exist only
when the right-hand side can be negative, that is when
the absolute value of ${\bf w}$ exceeds $w_{c0}$ defined by
\begin{equation}
w_{c0}^2 =\frac{2(\rho^3 g\sigma)^{1/2}}{\rho_{\rm_n}\rho_{\rm s}}\;,
                                                  \label{wc0}
\end{equation}
the instability taking place at ${\bf q}=\pm ({\bf w}/w)q_c$,
where $q_c^2={\rho g}/{\sigma}$.

On the other hand, for any finite $\eta>0$ one of the roots of Eq.
(\ref{dr}) crosses the real axis already when
\begin{equation}
S({\bf q})\equiv gq+\frac{\sigma}{\rho} q^3
                   -\frac{\rho_{\rm s}}{\rho}({\bf wq})^2
\end{equation}
touches zero, that is at
\begin{equation}
|{\bf w}|=w_c\equiv
\left[\frac{2(\rho g\sigma)^{1/2}}{\rho_{\rm s}}\right]^{1/2}
=\left(\frac{\rho_{\rm n}}{\rho}\right)^{1/2}w_{c0}\;,
                                                     \label{wc}
\end{equation}
the instability appearing at the {\em lower} value of relative velocity
than in absence of dissipation, although at the same value of ${q}$.
Note that in the limit of zero temperature (when
$\rho_{\rm s}\rightarrow\rho$) the criterion (\ref{wc}) coinsides
with the Landau criterion for creation of ripplons
in the reference frame of container walls.

For $S({\bf q})$ sufficiently close to zero
the value of the root which crosses real axis is given by
\begin{equation}
\omega({\bf q})- {\bf v}_{\rm n}^0{\bf q}\approx
\frac{1}{2}\frac{\rho S({\bf q})}{\rho_{\rm s}{\bf wq}+i\eta q^2}\;.
                                                        \label{wq}
\end{equation}
This shows that for small viscosity and $w$ just above $w_c$ the rate of
the development of the instability decreases with decreasing $\eta$
contrary to what is natural to expect.

By looking when the free energy of a corrugation, calculated
in the reference frame of the normal component, is no longer positively
defined (such approach can be considered as a macroscopic
generalization of the Landau criterion), the threshold for
the instability of the interface between two different
superfluids has been found \cite{V} to be given by
\begin{equation}
\rho_{\rm s1}({\bf v}_{\rm s1}^0-{\bf v}_{\rm n}^0)^2+
\rho_{\rm s2}({\bf v}_{\rm s2}^0-{\bf v}_{\rm n}^0)^2=2(F\sigma)^{1/2}\;,
                                                             \label{AB}
\end{equation}
where $F$ is a generalized restoring force, the role of which in
the case of free surface is played by $\rho g$.
In the limit when the density of one of the liquids goes to zero,
Eq. (\ref{AB}) is reduced to our criterion (\ref{wc}) obtained
for a free surface of a single superfluid liquid.

\section{Conclusion}

In the present work we have investigated the dynamic instability
of a superfluid liquid free surface caused by the relative motion of
superfluid and normal components along the surface.
The value of the instabilty threshold for finite viscosity,
given by Eq. (\ref{wc}), has been found to be independent of viscosity,
but lower than in absence of dissipation.
The same criterion can be obtained by looking for the thermodynamic
instability in the reference frame of the normal component
(which in equilibrium is fixed by container walls).

Analogous modification  of the instability threshold has been found
\cite{V} to take place on the interface between two superfluids
in presence of a friction with respect to the reference frame related
with container walls \cite{Kagan2}, %\footnote[2]
which leads to violation of the Galilean invariance.
Note that in our problem the same phenomenon appears in situation
when the form of dissipation does not imply the explicit selection
of the particular reference frame.
Nonetheless, the presence of dissipation (a finite value of viscosity)
turnes out to be sufficient to produce the same criterion
for surface instability as in the case when the form of the friction
leads to the direct violation of the Galilean invariance.

The first experimental observation of the Kelvin-Helmholtz instability
on the interface between $^3$He-A and $^3$He-B
by Blaauwgeers {\em et al.} \cite{BEV} have unambigously demonstrated
that it indeed takes place not at the classical, but at the modified value
of the threshold. According to our results, the same can be expected from
the instability on free surface of superfluid $^4$He.

$~$

The author is grateful to G.E. Volovik for useful discussion.
This work has been supported by the Program "Quantum Macrophysics"
of the Russian Academy of Sciences, by the Program "Scientific
Schools of the Russian Federation" (grant No. 00-15-96747), by the
Swiss National Science Foundation and by the Netherlands Organization
for Scientific Research (NWO) in the framework of Russian-Dutch
Cooperation Program.

% \end{multicols}
  
\end{multicols}


\begin{thebibliography}{99}
\bibitem{Kagan2}{The same type of dissipation has been taken into account
                by Kagan \cite{Kagan} when studying the instability of
                the quantum interface between superfliud and solid $^4$He.},
\end{thebibliography}
\end{document}